# The collapse of linear polyelectrolyte chains in a poor solvent: When does a collapsing polyelectrolyte collect its counter ions?

*Peter Loh, G. Roshan Deen[#], Doris Vollmer[§], Karl Fischer, Manfred Schmidt[\*], Arindam Kundagrami[&],*

*Murugappan Muthukumar[&]*

Institute of Physical Chemistry, University of Mainz, Welder Weg 11, 55099 Mainz, Germany

[#]Present address: Department of Chemistry and the Nano Science Center, University of Aarhus, Langelandsgade 140, Aarhus, Denmark.

[§]Max-Planck-Institute for Polymer Research, Ackermannweg 10, D-55118 Mainz, Germany

[&]Department of Polymer Science and Engineering, Conte Research Center, University of Massachusetts, 120 Governors Drive, Amherst, MA 01003, USA

mschmidt@uni-mainz.de

ABSTRACT: In order to better understand the collapse of polyions in poor solvent conditions the effective charge and the solvent quality of the hypothetically uncharged polymer backbone need to be known. In the present work this is achieved by utilizing poly-2-vinylpyridine quaternized to 4.3% with ethylbromide. Conductivity and light scattering measurements were utilized to study the polyion collapse in isorefractive solvent/non-solvent mixtures consisting of 1-propanol and 2-pentanone, respectively, at nearly constant dielectric constant. The solvent quality of the uncharged polyion could be quantified which, for the first time, allowed the experimental investigation of the effect of the electrostatic interaction prior and during polyion collapse, by comparing to a newly developed theory. Although the Manning parameter for the investigated system is as low as $l_B/l = 0.6$ ($l_B$ the Bjerrum length and $l$ the mean contour distance between two charges), i.e. no counterion binding should occur, a qualitative interpretation of the conductivity data revealed that the polyion chain already collects its counter ions when the dimensions start to shrink below the good solvent limit but are still well above the $\theta$-dimension.

**Introduction**

The conformation of linear polyelectrolytes in a good solvent is frequently investigated but the results are still controversially discussed, despite some progress in analytical theory as well as in computer simulation.[1,2] Problems originate from the subtle interplay between electrostatic interaction, intrinsic excluded volume and hydrophobic effects as well as from frequently ignored contributions of specific counter-ion and co-ion properties,[2,3] empirically expressed in the Hofmeister series.[4]

As compared to polyelectrolytes in a good solvent, the collapse of polyions in a poor solvent is much less investigated because of the further increase of complexity. Starting from the theoretical work of Khokhlov as early as 1980[5-9], there have been several papers based on theory[10-12] and computer simulations.[13-15] In particular the postulation of the "string of spheres" conformation by Dobrynin and Rubinstein[10] has been intriguing. Although few experimental studies on the collapse of polyelectrolytes have been published so far which seem to be compatible with the "string of spheres" picture, a solid experimental proof is still missing. [16-23]

In the present communication the chain dimension and the effective charge density of a slightly charged, high molar mass linear polyelectrolyte chain is investigated by light scattering and conductivity. By utilizing a mixture of a solvent and a non-solvent the experiments cover the whole regime from expanded coils to the collapsed state.

**Theory**

When a flexible polyelectrolyte chain is present in a solution containing its counterions and dissociated electrolyte ions, the effective charge of the polymer is modulated by adsorption of counterions. The extent of this adsorption is controlled by an optimization between attractive energies associated with the formation of ion pairs (among polymer segments and counterions) and the loss in translational entropy of the counterions, which would otherwise be free to explore the whole solution. Further, the distribution of counterions around the polymer depends on the polymer conformations, which in turn depend on how many counterions are adsorbing on the polymer molecule. Thus a self-consistent procedure is necessitated to calculate the size and the effective charge of the polyelectrolyte molecule.

Several computer simulations[14,15,24-28], where counterions are accounted for explicitly, clearly demonstrate that the effective charge indeed depends uniquely on the polymer size. Recently, an analytical theory has been presented, where the coupling among polyelectrolyte conformations, counterion adsorption, and translational entropy and electrostatic correlations of small ions was treated self-consistently.[30] Previously, only the progressive accumulation of counterions as the polymer coil continues to shrink due to an increasing number of ion-pair formation has been addressed. In the present paper, we extend this theory for the self-consistent determination of counterion accumulation as the polymer collapses due to hydrophobic forces.

Within the same context, there has recently been another model[8,9] where the polyelectrolyte chain and the background solution is treated as a three-state system. In this model, the unadsorbed counterions are presumed to partition into two domains, with one domain being the volume within the coil and the other domain outside the coil. In addition, the entropic part of chain fluctuations is not explicitly accounted for. The major prediction of the three-state model is that the polymer collapse induced by intra-chain hydrophobic attraction occurs in two stages with the accompanying two levels of counterion condensation. In contrast, in our theory, there is only one collapsed state and the extent of counterion adsorption follows the shrinkage of the polymer size. Here, we give only the key steps of the theory and the reader is referred to the original paper for more details.

The size and shape of a polyelectrolyte molecule in solutions are controlled by the binding energy of counterions onto the chain backbone, the translational entropy of counterions in the solution, and the concomitant changes in polymer conformations. The complete free energy $F$ of the system consisting of a flexible polyelectrolyte chain, its counterions and the added salt ions in high dilution is a sum of various contributions ($F = \sum_{i=1}^{5} F_i$)[30], and the components are given as follows. $F_1$, related to the entropy of mobility of the condensed counterions on the chain backbone, is given by

$$\frac{F_1}{Nk_BT} = (f_m - \alpha_1)\log\left(1 - \frac{\alpha_1}{f_m}\right) + \alpha_1 \log\left(\frac{\alpha_1}{f_m}\right), \tag{1}$$

where $N$, $k_B$, $T$, $f_m$, and $\alpha_1$ are, respectively, the number of Kuhn segments in the chain, the Boltzmann constant, the absolute temperature, the number of ionizable groups per Kuhn segment, and the number of ionizable groups with condensed counterions per Kuhn segment. The ionizable groups are assumed to be uniformly distributed along the chain backbone. $F_2$, related to the translational entropy of the uncondensed counterions and the salt ions (i.e., the free and mobile ions in the solution), is given by

$$\frac{F_2}{Nk_BT} = \left(f_m - \alpha_1 + \frac{\tilde{c}_s}{\tilde{c}_p}\right)\log\left\{\tilde{c}_p\left(f_m - \alpha_1\right) + \tilde{c}_s\right\} + \frac{\tilde{c}_s}{\tilde{c}_p}\log\tilde{c}_s - \left(f_m - \alpha_1 + \frac{2\tilde{c}_s}{\tilde{c}_p}\right), \tag{2}$$

where $\tilde{c}_p = c_p l_0^3$ and $\tilde{c}_s = c_s l_0^3$ are, respectively, the dimensionless number densities of the monomers and the monovalent salt ions, and $l_0$ is the Kuhn segment length. $F_3$, related to the fluctuations arising from the electrostatic interactions among all mobile ions, is given by the Debye-Hückel form,

$$\frac{F_3}{Nk_BT} = -\frac{2}{3}\sqrt{\pi}\tilde{l}_B^{3/2}\frac{1}{\tilde{c}_p}\left\{\tilde{c}_p\left(f_m - \alpha_1\right) + 2\tilde{c}_s\right\}^{3/2}, \tag{3}$$

where $\tilde{l}_B = e^2/4\pi\varepsilon\varepsilon_0 k_B T l_0$ is the dimensionless Bjerrum length, with $e$, $\varepsilon$, and $\varepsilon_0$ being, respectively, the electron charge, the solvent dielectric constant, and the vacuum permittivity. $F_4$, the adsorption energy gain due to ion-pairs formation with condensation, is given by

$$\frac{F_4}{Nk_BT} = -\tilde{l}_B\delta\alpha_1, \tag{4}$$

where $\delta = C\varepsilon l_0$ with $C = 1/\varepsilon_l d$ as a system specific parameter. Here $\varepsilon_l$ and $d$ are, respectively, the 'local' dielectric constant related to the material of the polymer backbone and the dipole length of the ion-pairs. Finally, the free energy of the polymer chain considers the interaction energy $V(r)$ between the Kuhn segments separated by a distance $r$, where

$$\frac{V(r)}{k_BT} = w\delta(r) + f^2 l_B \frac{e^{-\kappa r}}{r} \tag{5}$$

is valid for monovalent ionic groups. Here $w$, $f = f_m - \alpha_1$, and $\kappa = \sqrt{4\pi l_B\left\{c_p\left(f_m - \alpha_1\right) + 2c_s\right\}}$ are, respectively, the strength of the excluded volume interaction, the degree of ionization, and the inverse Debye length. The first and second terms of the potential represent the excluded volume (non-electrostatic) and electrostatic interactions, respectively. Using a well-known variational method[29,30] $F_5$, the chain free energy, is obtained as

$$\frac{F_5}{Nk_BT} = \frac{3}{2N}\left(\tilde{l}_1 - 1 - \log\tilde{l}_1\right) + \frac{4}{3}\left(\frac{3}{2\pi}\right)^{3/2}\frac{w}{\sqrt{N}}\frac{1}{\tilde{l}_1^{3/2}} + \frac{w_3}{N\tilde{l}_1^3} + 2\sqrt{\frac{6}{\pi}}f^2\tilde{l}_B\frac{N^{1/2}}{\tilde{l}_1^{1/2}}\Theta_0(a), \tag{6}$$

where, $\tilde{l}_1 = 6R_g^2/Nl_0^2$, $w = \frac{1}{2} - \chi$, and $w_3$ are, respectively, the effective expansion factor, the excluded volume parameter, and a three-body interaction parameter, with $R_g$ and $\chi$ being, respectively, the radius of gyration and the Flory-Huggins chemical mismatch parameter. Here

$$\Theta_0(a) = \frac{\sqrt{\pi}}{2}\left(\frac{2}{a^{5/2}} - \frac{1}{a^{3/2}}\right)\exp(a)\,\mathrm{erfc}\left(\sqrt{a}\right) + \frac{1}{3a} + \frac{2}{a^2} - \frac{\sqrt{\pi}}{a^{5/2}} - \frac{\sqrt{\pi}}{2a^{3/2}} \qquad (7)$$

is a cross-over function with $a = \tilde{\kappa}^2 N\tilde{l}_1/6$, where $\tilde{\kappa} = \kappa l_0$ is a dimensionless inverse Debye length. The original free energy[30] contained an additional contribution from the attractive interactions between dipoles formed on the chain backbone. These dipolar interactions are found to modify the excluded volume parameter marginally for low degrees of chain ionizability, and hence are ignored in the analysis of the experiments relevant to this article. A major assumption to derive Eq. (6) is that of uniform swelling of the chain with spherical symmetry. Note that a positive and non-zero $w_3$ is required to stabilize the free energy in the case of a chain collapse below the Gaussian dimension ($\tilde{l}_1 = 1$) for negative values of $w$. The optimum radius of gyration $R_g$ and the degree of ionization $f$ of an isolated chain are derived by simultaneous minimization of the free energy of the system (chain, counterions, salt ions, and the solvent) with respect to $R_g$ and $f$. Numerically, $F$ is minimized self-consistently with respect to $f$ and $\tilde{l}_1$ to obtain the equilibrium values of the respective quantities in specific physical conditions stipulated by $T$, $\varepsilon$, and $\tilde{c}_s$. It must be remarked that the free energy described above is for a single polyelectrolyte chain in a dilute solution, and it is valid for concentrations of salt not too high (so that $\kappa^{-1} \geq l_B$ or $c_s \leq (8\pi l_B^3)^{-1}$ for a monovalent salt). It is valid, however, for all temperatures, and for any degree of ionization (or ionizability) of the polymer. In addition, the above free energy is equally applicable for multi-chain systems in infinitely dilute solutions in which the chains have negligible inter-chain interaction (either excluded volume or electrostatic). Qualitative analysis of the free energy shows that the size and charge of the polyelectrolyte chain are primarily determined by the energy gain of ion-pairs [which is linearly proportional to an effective Coulomb strength ($\tilde{l}_B\delta$)] relative to the translational entropy of the mobile ions in the expanded state and by the relative strength of $w$ to $w_3$ in the collapsed state. The parameter C in Eq. (4) is the only adjustable parameter taken to fit the experimental data. One notes that calibration of $w$ by the respective uncharged chain is necessary to eliminate the uncertainty in determining the non-electrostatic interactions in charged polymers and that can be performed by setting $w_3 = 0 = f$ in Eq. (6). Minimizing $F_5$ with respect to $\tilde{l}_1$, in this case, yields the familiar formula for chain expansion

$$\alpha^5 - \alpha^3 = \frac{4}{3}\left(\frac{3}{2\pi}\right)^{3/2} w\sqrt{N} \quad (8)$$

where $\alpha^2 = \tilde{l}_1$. The functional dependence of the non-electrostatic parameter $w$ on the solvent composition is established first by using Eq. (8) to determine $w$ from the expansion factor of the uncharged polymer chain. Later, by performing the double minimization of the free energy of the charged polymer, $\alpha$ and $f$ were determined with C as a parameter, and then compared with the experimental data.

**Experimental**

A high molar mass linear poly-2-vinylpyridine chain (PVP, molar mass $M_w = 8.5 \times 10^5$ g/mol, $M_w/M_n$ = 1.18, degree of polymerization $P_w$ = 8095, purchased from and specified by PSS, Mainz, Germany), was quaternized with ethyl bromide in nitromethane at 60°C to a degree of quaternization of 4.3% (QPVP$_{4.3}$, calculated $M_{w,p} = 8.9 \times 10^5$ g/mol, number average of chemical charges per chain 295).

It is assumed that the quaternization occurs randomly along the chain which leads to a Gaussian distribution of the number of charges per chain at constant chain length.

Light scattering and conductivity measurements were conducted in a mixture of 1-propanol and 2-pentanone at high dilution ($c_p \leq 12$ mg/L). Whereas 1-propanol represents a good solvent for uncharged PVP and QPVP$_{4.3}$, 2-pentanone is a non-solvent for both. The solvents are essentially isorefractive ($n_D$ (1-propanol) = 1.385, $n_D$ (2-pentanone) = 1.390). Thus complications in the interpretation of the light scattering data arising from preferential absorption are avoided.

The dielectric constants of the solvent mixtures range from $\varepsilon = 21$ (1-propanol) to $\varepsilon = 16$ (2-pentanone), and accordingly, the Bjerrum length varies from 2.7 nm to 3.5 nm for 0% and 100% 2-pentanone content, respectively (see Supporting Information). Given the contour distance between two charges along the chain, $b$ = 5.8 nm, no counter-ion condensation is expected to occur for the present system according to Manning.[31,32] In order to avoid a structure peak to be observable by light scattering, tetrabutylamonium bromide (TBAB) was added to the solution ($c_s$ = $10^{-5}$ mol/L) which causes the Debye screening length to range from 43 nm < $l_D$ < 49 nm depending on the solvent composition. At this added salt and polyion concentrations no "slow mode"[1,33-35] could be detected. Thus, the amount of added salt was chosen such that the known problems arising from intermolecular electrostatic interaction are minimized and that intramolecular electrostatic interactions are kept as large as possible. Given the above experimental conditions, the centers of mass of the polyions are separated by a mean distance of 300 nm. If all $f_mN$ charges were located at the center of mass, interaction energy of approximately 0.4 $k_BT$ would result between a pair of chains, according to the Debye-Hückel formula.

**Results and discussion**

The polyion conformation characterized by the square root of the apparent mean square radius of gyration, $R_g^{app} \equiv \langle R_g^{2\,app} \rangle_z^{1/2}$, the apparent hydrodynamic radius, $R_h^{app} \equiv \langle 1/R_h^{app} \rangle_z^{-1}$, and the apparent molecular weight $M_w^{app}$ are shown in Fig. 1 as functions of the weight fraction of the non-solvent 2-pentanone, $w_{ns}$. Although the concentration of the measured solution is extremely small ($c_p$ = 12 mg/L), long range electrostatic interactions due to only a small amount of screening salt may influence both, dynamic (DLS) and static (SLS) light scattering measurements as indicated by the superscript "app".

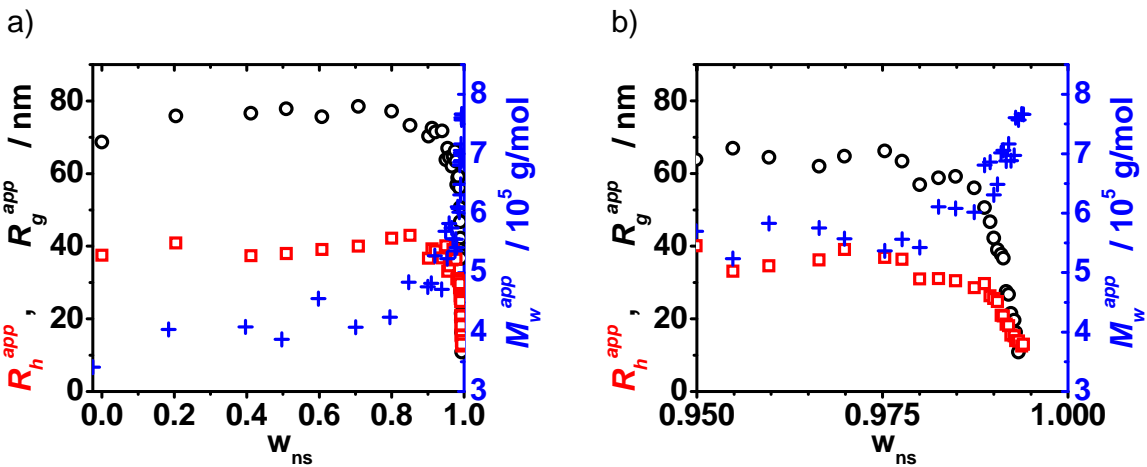

**Figure 1.** (a) Apparent radius of gyration, $R_g^{app}$, (circles, left axis), the apparent hydrodynamic radius $R_h^{app}$, (squares, left axis) and the apparent molar mass, $M_w^{app}$, (crosses, right axis) are plotted as functions of the weight fraction of non-solvent, $w_{ns}$. (b) Magnification of the collapse regime, same symbols as in (a).

The apparent molar mass at $w_{ns}$ < 0.8 is observed to be less than half of the true molar mass. Chain degradation during the quaternization reaction can be safely ruled out, because for larger amounts of added salt ($c_s = 10^{-4}$ M and $c_s = 0.1$ M) the true molar mass is obtained within experimental error). Alternatively, one might misinterpret the increase of the apparent molar mass of the polyelectrolyte in the collapse regime in terms of a true increase of molar mass caused by inter chain aggregation. However aggregation can be safely ruled out because i) the apparent molar mass for all $w_{ns}$-values is smaller than the true molar mass, ii) the increasing molar mass in the collapse regime approaches the true molar mass and iii) the correlation functions measured in the collapse regime do not exhibit broader relaxation time distributions than those at small 2-pentanone content.

Uncertainties in the determination of the refractive index increments, $dn/dc_p$, are also not likely to influence $M_w^{app}$ as discussed in more detail in the Supporting Information.

Rather, the disparity between the apparent molar mass $M_w^{app}$ and the true molar mass $M_{w,p}$ is governed by the dissociated counterions and by the osmotic coefficient $\Phi$ defined as[36,37]

$$\Phi = \pi / \pi^{id} = M^{id} / M^{app} \tag{9}$$

with $\pi$ and $\pi^{id}$ the measured and ideal osmotic pressures. For most polyion solutions the ideal osmotic pressure in volume $V$ is dominated by the very many dissociated counterions

$$\pi^{id} = RTn_p\left(1+fN\right)/V \tag{10}$$

with $n_p$ being the number of polyions. Since $fN \gg 1$, the ideal osmotic pressure does not yield the true polyion molar mass but represents a measure for counterion dissociation according to

$$M^{id} = M_{n,p}/\left(1+fN\right) \approx M_{n,p}/fN \tag{11}$$

In practice, eq. 11 cannot be utilized for the determination of counterion dissociation due to the non ideal behavior leading to the osmotic coefficient in eq. 9. However, since the scattering intensity extrapolated to $q = 0$ is inversely proportional to the osmotic compressibility, the ratio $M_{w,p}/M_w^{app}$ is proportional to the product $(\Phi\gamma)_{LS}$ according to

$$\left(\Phi\gamma\right)_{LS} = M_{w,p}/\left(M_w^{app} fN\right) \tag{12}$$

with $\gamma$ the fraction of dissociated counterions $f/f_m$. According to eq. 12 $M_w^{app}$ shown in Fig.1 is inversely related to the number of dissociated counterions. Consequently the effective charge density is observed to decrease with decreasing solvent quality which is discussed below.

It is to be noted that the measured $R_g$- and $R_h$-values shown in Fig. 1 may be falsified by an intermolecular structure factor and are marked as "apparent" quantities, accordingly. The $R_g^{app}$-values shown above were derived from the slopes of the reduced scattering intensity versus $q^2$ at $c_p = 12$ mg/L, i.e. without extrapolation to infinite dilution. The slopes vs. $q^2$ were strictly linear for all solvent compositions (see Supporting Information for some examples). However, the experimentally observed linearity does not necessarily prove the slope to be unaffected by intermolecular interference effects as pointed out in the literature.[38,39] The hydrodynamic radius may be affected by the static structure factor as well.[1] Without going into detail intermolecular electrostatic interactions should always yield smaller $R_g^{app}$ and $R_h^{app}$-values as compared to interaction-free values, the effect on $R_h$ being more pronounced than on $R_g$.[1,38-41] As will be shown below the electrostatic interaction is significantly reduced in the collapse regime which constitutes an additional complication. A quantitative discussion of concentration effects on the experimentally observed dimensions will be presented in a future publication.

Keeping these uncertainties in mind, the measured $R_g$- as well as the $R_h$-values remain constant for $0 < w_{ns} < 0.8$, followed by a slight decrease until for $w_{ns} \approx 0.988$, above which the polyelectrolyte chains collapse and eventually aggregate and phase separate at $w_{ns} > 0.994$. The regime close to the phase boundary, $0.95 < w_{ns} < 1$, is enlarged in Fig. 1b for better clarity.

In order to quantify the solvent quality of the uncharged polyvinylpyridine the expansion factor $\alpha = R_g/R_{g,\theta}$ and the second virial coefficient, $A_2$, were measured as functions of $w_{ns}$ (see Fig. 2 and 3) and compared to the theory (Eq. (8)) yielding the solvent quality parameter $w$ as a function of $w_{ns}$. Note that Eq. (8) is only valid for $w \geq 0$. For $w < 0$, we used values of $w$ linearly extrapolated from its value at the $\theta$-condition. $\theta$-dimensions ($A_2 = 0$, $R_{g,\theta} = 29$ nm) were observed at $w_{ns} = 0.91$ (see Fig. 2 and 3). Below theta dimensions ternary interactions were included (the third term in Eq. (6)). As mentioned in the theory, a non-zero positive value of $w_3$ was required to stabilize the chain collapse below $\theta$-dimensions. A fixed value of $w_3 = 0.00165$ was chosen for both the uncharged and charged polymers used in our experiments.

In Fig. 4 the expansion factor, $\alpha = R_g^{app}/R_{g,\theta}$, of the uncharged PVP chains (a) and of the polyions (b) is plotted versus $w_{ns}$ along with the theoretical prediction. Quantitative agreement is observed except close to the phase transition where the experimental data show a broader phase transition regime as discussed in some detail, below.

Using the same dependency of $w$ on $w_{ns}$, the expansion factor $\alpha$ for the charged chain is fitted by the theoretical curve with only one adjustable parameter, $C = 0.183$nm$^{-1}$, which reflects on the local dielectric constant in the vicinity of the polyion backbone, $\varepsilon_l$. Since the ion-pair energy ($\tilde{l}_B \delta$) and the temperature remain constant for the entire range of the experiment, the effective charge $f$ has negligible variation in the expanded state. The chain free energy [Eq. (6)] in this state is dominated by the electrostatic term, and consequently the theoretical chain dimension (corresponding to an approximately constant charge density) varies little. Nevertheless, one notes that with increasing proportions of $w_{ns}$ there is a slight increase in chain size due to a small increase in the value of Bjerrum length (with decreasing dielectric constant) that marginally enhances the intra-chain monomeric repulsion captured in the forth term of the free energy [Eq. (6)]. This small increase in the dimensions predicted by theory is smaller than the experimental uncertainty for the $R_g$ determination in the regime $0 < w_{ns} < 0.8$. However, the size and shape of the chain undergoes a drastic change at a threshold poorness of the solvent. Beyond the threshold poorness the chain collapses, and that leads it to collect its counterions.

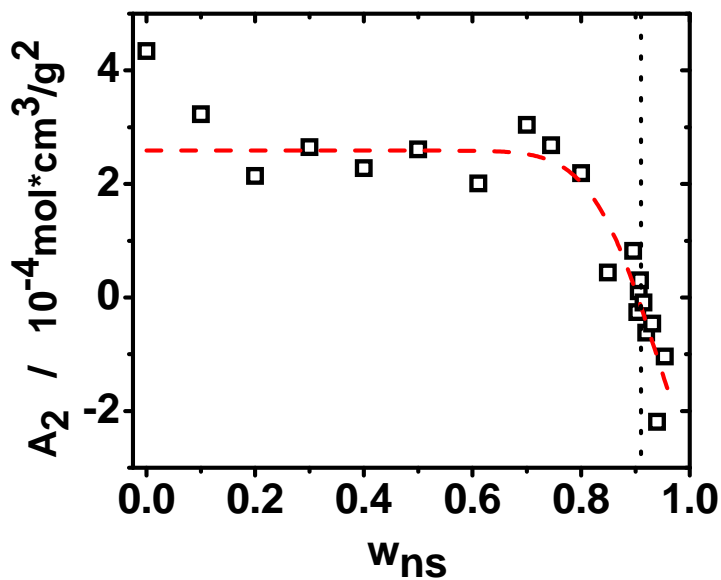

**Figure 2.** The second virial coefficient $A_2$ for the neutral polyvinylpyridine is plotted against the volume fraction of the non-solvent, $w_{ns}$ (the dashed red line serves as a guide to the eye). θ-conditions are marked by the dotted vertical line at $w_{ns} = 0.91$

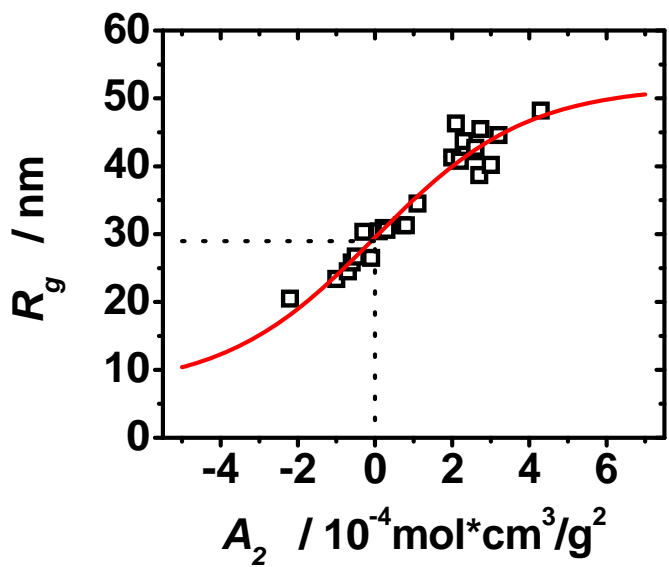

**Figure 3.** The radius of gyration for the neutral polyvinylpyridine is plotted versus its second virial coefficients, $A_2$, determined at different $w_{ns}$. A sigmoidal fit having its inflection point set to $A_2 = 0$ (red line) describes the data satisfactorily and yields the θ-dimensions marked by the two dotted lines ($A_2 = 0$, $R_{g,\theta} = 29$ nm).

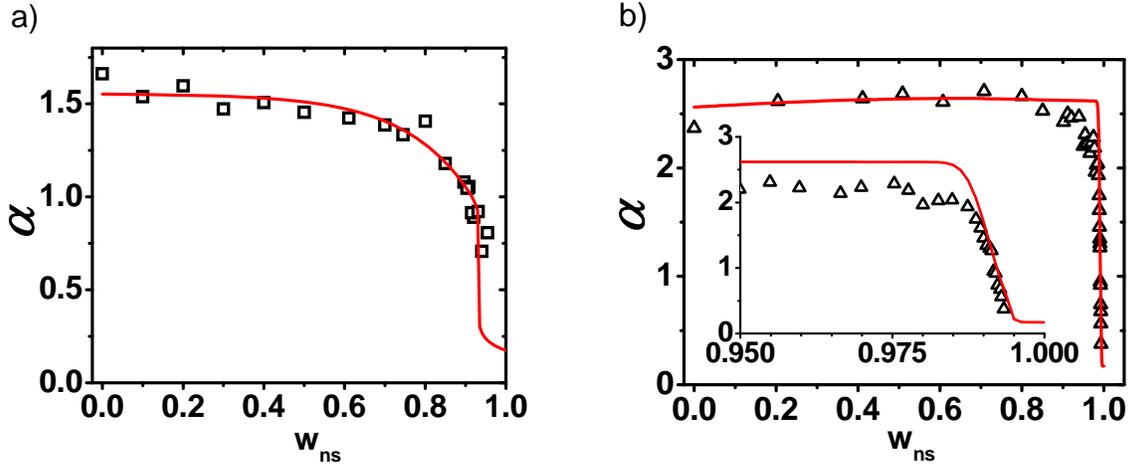

**Figure 4.** The expansion factor $\alpha$ is plotted against the volume fraction of the non-solvent, $w_{ns}$, for the neutral polyvinylpyridine (a) and for the charged QPVP$_{4.3}$ sample (b) The red lines represent the fit according to Eq. (8) with $f = f_m = 0$ and $w_3 = 0.00165$ (a) and by minimizing the five contributions to the free energy as described in the theoretical part with $C = 0.183$ nm$^{-1}$ as the only fit parameter.

It must be noted that the theory presented above predicts a first-order coil-globule transition for the chains if the excluded volume parameter $w$ is smaller than a certain threshold value, and provided that the three-body interaction parameter $w_3$ is also smaller than a critical value. The strength of the three-body interaction parameter pertinent to our analysis is substantially lower than the critical value, and hence the theory predicts a first-order phase transition sharper than the relatively broad transition regime observed in the experiments. One should note, that polydispersity in the chain length ($N$) and in the maximum degree of ionization ($f_m$) could broaden the transition due to a distribution of the threshold value of $w$. Whereas the chain length distribution has little effect (data not shown), the variation of the number of charges per chain at constant chain length assuming a Gaussian distribution was utilized for the fit shown in Fig 4b (for details see Supporting Information). So far, no explanation can be given for the experimentally observed small decrease of α in the regime $0.8 < w_{ns} < 0.985$.

The value of 0.183 nm$^{-1}$ for $C$ is equivalent to $\varepsilon_l$ = 10.9 if the dipole length is assumed to be 0.5 nm. This value of $\varepsilon_l$ is in between the dielectric constant ($\varepsilon$ = 8.33) of 2-ethylpyridine, which is chemically close to the chain backbone, and that of the solvent ($16 < \varepsilon < 21$). However, given the uncertainty in the dipole length $d$, i.e., the mean distance of the bound counterions from the respective polyion charges (which can vary from 0.3 nm to a few nm), the value for $\varepsilon_l$ given above should not be over interpreted. Nevertheless, we notice that the value of $C$ is remarkably close to the value 0.175 nm$^{-1}$ estimated for polymers of type sodium polystyrene sulfonate (NaPSS) in the original theory.[30]

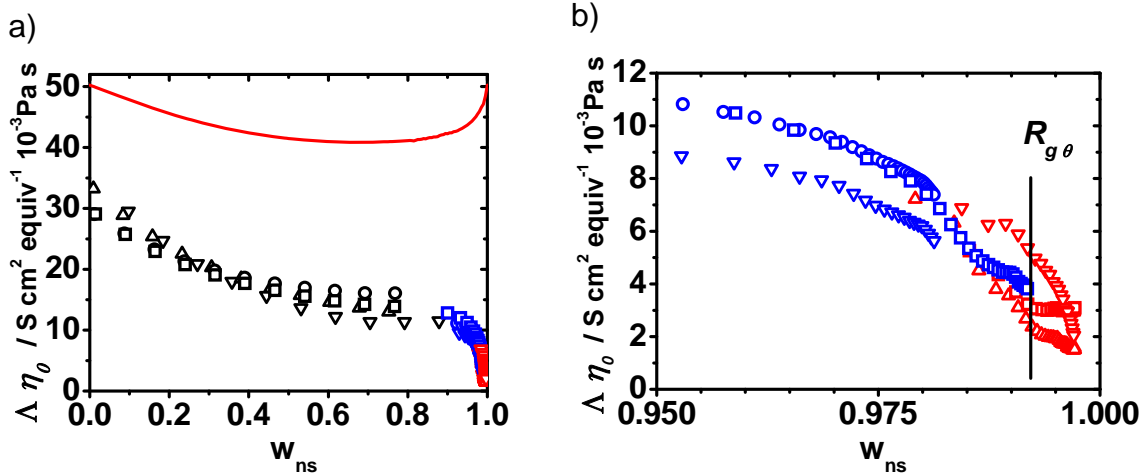

**Figure 5.** (a) Walden product of the bare salt solution (curve) and of the polyelectrolyte solution for different concentrations; squares: 12mg/L; circles: 10 mg/L; triangles up: 8 mg/L; triangles down 6 mg/L. The different colors indicate a different dilution series. (b) Magnification, symbols as in (a).

In Fig. 5 the equivalent conductivity corrected by the solvent viscosity, $\Lambda\eta_0$, the so-called Walden product is shown for the bare salt solution and of the polyion solutions at different concentrations of $6 < c_p < 12$ mg/L. For the bare salt solution the Walden product is expected to be independent of solvent composition; however, the $\Lambda\eta_0$ values for the TBAB depend on $w_{ns}$ as shown by the curve in Fig. 5 in qualitative agreement with literature data.[42-45] Specific ion solvation effects, microviscosity and liquid dynamics were postulated to cause the non-ideal behavior of the conductivity data.[43,46] For TBAB in mixtures of 1-propanol (forming hydrogen bonds) and acetone (solvating cations strongly) it was concluded[48] that preferential solvation of both the tetrabutylammonium-cation as well as the bromide cause the peculiar dependence of the Walden product on solvent composition.

Compared to the TBAB, the Walden products of the polyion solutions decrease in a similar fashion with increasing pentanone content but a bit more pronounced in the regime $0 < w_{ns} < 0.7$. However, for $w_{ns} > 0.9$ the Walden products of the polyion strongly decrease to very small values whereas the bare salt mobility increases slightly. The Walden product as a function of solvent composition is qualitatively similar to measurements of QPVP in mixtures of methanol/2-butanone[47] and of poly(methacryloylethyl trimethylammonium methylsulfate) in mixtures of water/acetone.[19] Obviously this strong decrease cannot be explained by subtle ion solvation effects as discussed above but rather reflects the association or binding of counter ions onto the polyion chain. It should be noted that the concentration dependence of the Walden product is very small which indicates that for the present conditions the conductivity of the solvent is not significantly influenced by the polyions and that interionic dynamic coupling effects are small. Approaching the phase transition the polyion chain starts to collect and bind its counter-ions as the chain dimensions become successively smaller. Eventually, the collapsed polyion chain preserves a few charges only, most probably some surface charges known from colloids. This experimental observation is in remarkable qualitative agreement with the results of explicit solvent simulations.[15] Interestingly, the polyion mobility is already significantly reduced well before the unperturbed $\theta$-dimension is reached (see Fig. 2 and 3). The obvious strong charge reduction in a regime where the Bjerrum length changes by 5% only, questions the applicability of the Manning condensation concept[31,32] to flexible polyelectrolyte chains at least for poor solvent conditions.

Following the general conductivity theory based on non-equilibrium thermodynamics,[48-52] and ignoring interionic friction effects, the electrolytic conductivity of a polyelectrolyte solution in the presence of added salt, $\sigma$, is given by

$$\sigma = \sigma_s + \gamma \left( \lambda_{Br} + \lambda_{Poly} \right) c_p' \tag{13}$$

where $\sigma_s$ represents the conductivity of the bare salt solution (in S/cm), $\lambda_{Br}$ and $\lambda_{Poly}$ are the electrophoretic mobilities (in S cm$^2$/equiv) of the bromide ion and the polyion, respectively and $c_p'$ is the equivalent concentration of the polyelectrolyte (in equiv/L). The equivalent conductivity of a polyelectrolyte solution, $\Lambda = (\sigma - \sigma_s) / c_p'$ may then be expressed as

$$\Lambda = \gamma \left( \lambda_{Br} + \lambda_{Poly} \right) \tag{14}$$

Ignoring all dynamic coupling and screening effects between polyions and counterions/salt ions the fraction $\gamma$ may be determined from $\Lambda$ by eq. 14 with the known mobility of the bromide ions and the following simplified expression for the polyion mobility

$$\lambda_{Poly} = \gamma Z F e / \left( 6 \pi \eta_0 R_h^{app} \right) \tag{15}$$

with the Faraday constant F, the solvent viscosity $\eta_0$ ( see Supporting Information), and $Z$ the number of chemical charges per chain, $Z = f_m N$.

In Fig. 6 the fraction of the effective charges $\gamma = f / f_m$ derived from the conductivity data by eqs. 14 and 15 is compared to the theoretically predicted charge density obtained through the double minimization of the free energy [Eq. (1-6)]. The observed qualitative agreement was to be expected in view of the perfect match of the expansion factor shown above. As mentioned before, the theoretical charge density in the expanded state is found to be virtually constant due to the absence of variation in the effective Coulomb strength $\tilde{l}_B \delta$. Note that this happens despite the somewhat decreasing value of the bulk dielectric constant with increasing $w_{ns}$, because the Coulomb strength relevant to the ion-pair energy depends only on the local (not the bulk) dielectric constant related to the material of the polymer backbone. Again the experimental data show a broader phase transition regime, but the location of the phase boundary where no free counterions exist is well reproduced. The theory predicts a first-order coil-globule transition for both, size and effective charge of the polymer chain. For comparison, also the product $(\Phi\gamma)_{LS}$ of the osmotic coefficient $\Phi$ and the fraction of dissociated counterions $\gamma$ is shown. It qualitatively compares well to $\gamma$ derived from the conductivity measurements. However, the osmotic coefficient is in the order of 0.01, a clear indication for highly non-ideal solution behavior, the origin of which will be investigated in some more detail in future work.

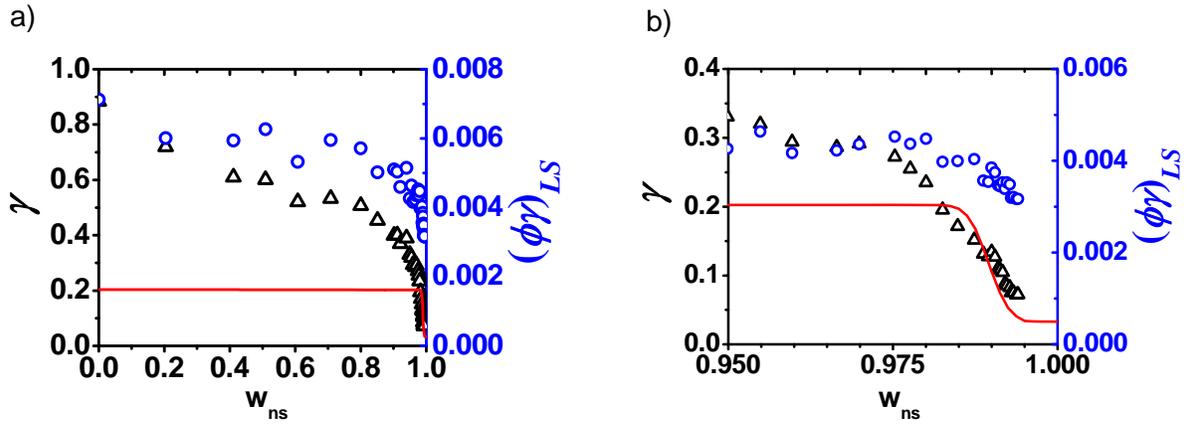

**Figure 6.** (a) Effective charge density $\gamma = f/f_m$ (triangles, left scale) and the osmotic coefficient $(\Phi\gamma)_{LS}$ (circles, right scale) as a function of the non-solvent fraction, $w_{ns}$. The solid curve shows the theoretical charge density $\gamma = f/f_m$. (b) Magnification, symbols as in (a).

Although being an apparent quantity only as discussed above, an interesting behavior of the ratio $R_g^{app}/R_h^{app}$ is observed as shown in Fig. 7. Despite some significant reduction in the absolute chain dimensions the ratio $R_g^{app}/R_h^{app}$ remains on a high level close to 2 which lies well above the theoretical limit of neutral flexible coils in the excluded volume limit, $R_g/R_h = 1.73$. Only for $w_{ns} > 0.993$ the ratio reduces to smaller values around 1 as expected for spherically collapsed coils. A similar behavior was also observed for the $Ca^{2+}$ and $Cu^{2+}$ induced collapse of polyacrylic acid[53] and of polymethacrylic acid.[54] For the $Ca^{2+}$ and $Sr^{2+}$ induced collapse of NaPA a string of sphere collapse was postulated by SANS[55] and by anomalous x-ray scattering experiments,[56] respectively. However, the divalent counterion induced collapse of polyelectrolytes has an entirely different physical origin as compared to the collapse of polyelectrolytes in a poor solvent.[57]

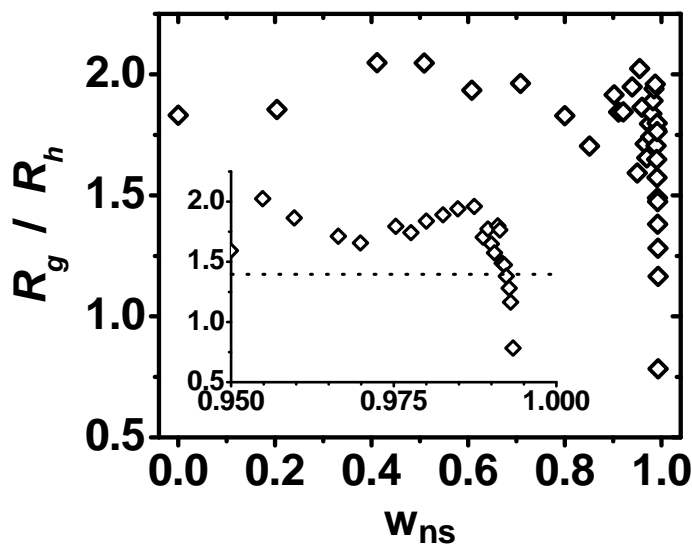

**Figure 7.** The apparent $\rho$-ratio $R_g^{app}/R_h^{app}$ is plotted vs. the fraction of non solvent $w_{ns}$. Inset: Magnification of the collapse transition with the dotted line indicating $R_g/R_h = 1.4$ experimentally found for the uncharged PVP ($\theta$-condition).

**Conclusion**

The combination of conductivity and light scattering measurements is well suited to investigate cooperative effects of counterion binding and chain collapse mediated by solvent quality and electrostatic interaction. Since the dielectric constant of the solvent remains virtually constant during the chain collapse, the counterion binding is entirely caused by the reduction in the polyion chain dimension. Remarkably the counterion binding occurs already well above the theta dimension of the polyion which was also reported for the $Sr^{2+}$ induced collapse of sodium polyacrylate (NaPA) in aqueous sodium chloride solution.[58] The theory of uniform collapse induced by concomitant counterion binding agrees quantitatively with the location of the phase boundary, but does not properly reproduce the width of the transition as mentioned above. Besides possible anisotropic chain conformations specific ion-solvation effects could also be the origin of the observed discrepancy.

Future work will focus on the variation of the degree of quaternization as well as on the influence of the molar mass and of the chemical charge density, particularly at high charge.

**Acknowledgement.** The work was supported by the German Science Foundation (DFG grant SCHM 553/19-2), by the International Max Planck Research School "Polymers in Advanced Materials", Mainz (Stipend for P. L.) and by the Humboldt Foundation (Humboldt prize for M. M.).